\documentclass[epj]{svjour}
\usepackage{graphics}
\def\ohalf{{\textstyle{1\over 2}}}

\def\half{{\textstyle{1\over 2}}}
\def\vqhalf{{\textstyle{\vec{Q}\over 2}}}

\def\osix{{\textstyle{1\over 6}}}

\def\tthird{{\textstyle{2\over 3}}}
\def\fivehalf{{\textstyle{5\over 2}}}

\newcommand{\beq}{\begin{equation}}
\newcommand{\be}{\begin{equation}}
\newcommand{\eeq}{\end{equation}}
\newcommand{\ee}{\end{equation}}
\newcommand{\beqa}{\begin{eqnarray}}
\newcommand{\bea}{\begin{eqnarray}}
\newcommand{\eeqa}{\end{eqnarray}}
\newcommand{\eea}{\end{eqnarray}}
\newcommand{\bra}[1]{\langle {#1} |}                        
\newcommand{\ket}[1]{| {#1} \rangle}

\usepackage{graphicx}
\usepackage{bm,hyperref,color}
\usepackage{epsfig}
\begin{document}

\title{Electroweak properties of the $\pi$, $K$ and $K^*(892)$
in the three forms of relativistic kinematics}

\author{
Jun He\inst{1,2,3}
\thanks{Email address: hejun@ihep.ac.cn}
\and
B. Juli\'a-D\'{\i}az \inst{4,5}
\thanks{Email address: bjulia@dapnia.cea.fr}
\and
Yu-bing Dong \inst{1,2}
\thanks{Email address: dongyb@mail.ihep.ac.cn}
}                     
%
%
\institute{
CCAST(World Lab.), P.O.Box 8730, Beijing 100080
\and
Institute of High Energy Physics, Chinese Academy of Sciences,
P.O.Box 918-4, 100049 Beijing, P.R. China
\and
Graduate School of the Chinese Academy of Sciences, 100049, Beijing, P.R. China
\and
Department of Physics and Astronomy, University of Pittsburgh, PA 15260, USA
\and
Helsinki Institute of Physics and Department of Physical Sciences,
P.O.Box 64, 00014 University of Helsinki, Finland}
\date{Received: date / Revised version: date}
%
\abstract{
The electromagnetic form factors, charge radii and decay constants
of $\pi$, $K$ and $K^*(892)$ are calculated using the three forms
of relativistic kinematics: instant form, point form and (light)
front form. Simple representations of the mass operator together
with single quark currents are employed with all the forms. Making
use of previously fixed parameters, together with the constituent
quark mass for the strange quark, a reasonable reproduction of the
available data for form factors, charge radii and decay constants
of $\pi$, $\rho$, $K$ and $K^*(892)$ is obtained in front form.
With instant form a similar description, but with a systematic
underestimation of the vector meson decay constants is obtained
using two different sets of parameters, one for $\pi$ and $\rho$
and another one for $K$ and $K^*(892)$. Point form produces a poor
description of the data.
} 
\maketitle

\section{Introduction}

The understanding of electromagnetic and weak properties
of low mass hadrons is still an open issue, mostly due to the
fact that the theory of the strong interactions, QCD,
cannot be easily solved at low energies. This includes,
e.g., the description of the spectra of bound states of
quarks, baryons and mesons, and reactions involving the
excitation of resonances. These difficulties in solving
QCD in the nonperturbative regime have triggered many
investigations, more or less related to QCD, which try
to shed some light on this domain.

One of these approaches, which we explored here, is the
formulation of relativistic quark models with a fixed number of
degrees of freedom. Relativistic quark models have been
implemented in three ways, depending on the way in which the
interactions are included in the commutator relations of the
Poincar\'e algebra~\cite{dirac,keister}. In principle the three
ways should provide similar results. However, in practice, the use
of simplifications, notably the use of single quark currents which
permits a simpler picture of the process, forces the appearance of
qualitative differences in the results.

In a previous work~\cite{junhe} the electromagnetic form factors
of $\pi$ and $\rho$ were studied making use of three different
forms of relativistic quantum mechanics. The ground state wave
function was adjusted in each of the forms to describe both the
charge radii and the high-$Q^2$ behavior of the pion charge form
factor. It was found that front and instant forms permitted a
reasonable reproduction of the pion form factors, and a coherent
picture for the form factors and charge radii of $\rho$. With point
form no ground state could be found, within the considered wave
functions, such that the pion charge form factor would be
qualitatively reproduced.

The main purpose of this work is to explore to what extent the
mass operators which were fixed to reproduce $\pi$ form factors
in each of the forms and which were applied to the study of
its vector partner, $\rho$, are able to provide a description also
of the other members of the $SU(3)_f$ octet, in our case the $K$
and the vector $K^*(892)$.

Comparison with experimental data for the case of the kaon form
factor and decay constants and with some of the previous works
done in any of the three forms of kinematics are
given~\cite{Chung:mu,Cardarelli,ChoiKaon,Xiao,Krutov,Amghar:2003tx}.

The point form used in this work, which follows
Refs.~\cite{Wagenbrunn,Riska04,junhe}, differs from the one
discussed lately in Refs.~\cite{Amghar:2003tx,Desplanques02}
where a closer contact with the original Dirac formulation is
pursued. The formulation used here emphasizes the relevant fact
that distinguishes among the forms which is the kinematic
subgroup of the Poincar\'e group. Once a kinematic subgroup
(``form of kinematics'') is chosen, the main difference
between the forms of kinematics, when considering single
quark currents, lies in the way the variables entering in the rest
frame wave functions are related to the variables appearing in the
interaction vertex. This distinction between the two formulations
of the point form is not quantitatively very relevant for the case
of two-body systems as was shown in Ref.~\cite{Desplanques04}.

This article is organized in the following way: Section~\ref{sec:wf}
presents the wave functions used. Then Section~\ref{sec:ew}
contains the formulas needed to compute the form factors and
decay constants of both spin-0 and spin-1 mesons, including
mesons made up of quarks of different mass. The $\pi$ and the
$K$ are studied in Section~\ref{sec:pik}. The decay constants of
$\rho$ and $K^*(892)$ and the form factors of the $K^*(892)$
are presented in Section~\ref{sec:kst}. A summary and discussion
are given in the last section.

\section{Wave functions}

\label{sec:wf}

In the rest frame, meson states are represented by eigenfunctions
of the mass operator, which are functions of internal momenta,
$\vec k_i$, and spin variables. A simple spectral representation
of the mass operator, with meson wave functions constructed in the
naive quark model~\cite{Close}, is considered,
\begin{eqnarray}
\psi^{\pi(K)}({\vec q}) &=& \xi_c \, \varphi_{0}({\vec q}) \,
\phi_S \,\chi_A \,,\nonumber \\
\psi^{\rho(K^*)}({\vec q}) &=& \xi_c \, \varphi_{0}({\vec q}) \,
\phi_A \,\chi_S \, ,
\end{eqnarray}
where $\xi_c$, $\phi_{S}$ and $\chi_A$ are the color, flavor and
spin wave functions.

The effect of the Lorentz transformation on the spin variables for
canonical spins is accounted by a Wigner rotation of the form:
\beq
D^{1/2}_{\lambda_i,\sigma_i}\left(R_W[B(v_K),k_i]\right)
\eeq
with
\beq
R_W[B(v_K),k_i]:= B^{-1}(p_i)B(v_K)B(k_i)\, ,
\eeq
where $B(v)$ are rotationless Lorentz transformations, and $v_K$ is the boost
velocity.

For the spatial part of the wave function, both Gaussian and
rational forms are employed:
\beqa
\varphi^G_{0}({\vec q})&=&\frac{1}{(b\sqrt{\pi})^{3/2}} e^{-{\vec q}^{\,2}/2b^2}\,, \nonumber
\\
\varphi^{R}_{0}({\vec q})&=& {\cal N}(1+{\vec q}^{\,2}/2b^2)^{-a},
\label{eq:wf}
\eeqa
where ${\vec q}=\frac{1}{\sqrt{2}}({\vec
k}_2-{\vec k}_1)$ and ${\cal N}$ is a normalization constant. In
the center of mass frame we have ${\vec k}_1+{\vec k}_2=0$ and
thus $\vec k_2 =\frac{1}{\sqrt{2}} \vec q =-\vec k_1$. As a
starting point the parameters used in Ref.~\cite{junhe} which are
given in Table~\ref{parameters} are employed.

\begin{table}[t]
\begin{center}
\begin{tabular}{ l |r|r|c|r}
\hline \hline
                     &b [MeV]    &$m_q$[MeV]  & a       &  $m_s$[MeV]\\
\hline \hline
\multicolumn{5}{c}{Gaussian}\\
\hline
Instant form         &370 [470]  &  140 [200] &  $--$   &500\\
Point form           &3000 [470] &  380 [200] &  $--$   &500\\
Front form           &450  [500] &  250 [250] &  $--$   &400\\
\hline
\multicolumn{5}{c}{Rational}\\
\hline
Instant form         &700 [520]  &  150 [250] &5 [3]    &500\\
Point form           &3000 [520] &  300 [250] &1 [3]    &500\\
Front form           &600 [650]  &  250 [250] &3 [3]    &400\\
\hline
\end{tabular}
\caption{Parameters used in instant, point and front form both for
the rational and gaussian spatial wave functions. In brackets are
the readjusted sets of parameters as explained in the text in
Section~\ref{sec:pik}. \label{parameters}}
\end{center}
\end{table}

The Jacobian of the transformations between the variables are, for
point form,
\beq J(\vec v;\vec p_2):= \left({\partial \vec q \over\partial \vec p_2}\right)_{\vec v}
= 2\sqrt{2}{(E_2 v^0-p_{2z}v_z)\over E_2}\, , \label{pointjac}
\eeq
with
\bea
\omega_i=\sqrt{m_i^2+k_i^2} \, ,
\quad E_i=\sqrt{m_i^2+p_i^2} \, ,
\eea
for front form
\bea J({\bf P};{\bf p_2})&:=& \left({\partial\vec q \over \partial (\xi_2,{\bf k}_{2\perp})} \right)_{\bf P}
\nonumber \\
&=&2\sqrt{2}\frac{M_0}{4\xi(1-\xi)}\left[1-\left(\frac{m_1^2-m_2^2}{M_0^2}\right)^2\right]\,,
\eea
with
\beqa k_{zi}&=&{1\over2}\left(\xi_iM_0-{m_q^2+k_{i\perp}^2\over \xi_i M_0}\right) \,,
M_0^2= \sum_i{m_i^2+k_{i\perp}^2\over \xi_i},
\eeqa
and for instant form,
\beq
{\cal }J(\vec P,\vec p_2):=2\sqrt{2}\frac{\omega_2}{E_2}\left\{1-\frac{E_2v_z}{M_0}\left(
{p_{1z}\over E_1}-{p_{2z}\over E_2}\right )\right\},
\label{instjac}
\eeq
where
\beq
P_x=P_y=0\;, \; M_0^2= (\sum _i
E_i)^2-|\vec P|^2 \; , \; \vec v := {\vec P\over M_0}\,.
\eeq

\section{Meson electroweak properties}

\label{sec:ew}

As in Refs.~\cite{Chung:mu,Riska04} the effective
conserved electromagnetic current operator in
each of the forms can be generated by the dynamics
from a current which is covariant under the kinematic
subgroup. Then, electromagnetic form factors of
two-body systems can be defined as certain matrix
elements of the electromagnetic current. In point
and instant forms, the charge form factor of scalar
mesons can be defined as follows,
\begin{eqnarray}
F_C(Q^2)=\langle 0, {\vec Q}/2|I^0(0)|0,-{\vec Q}/2\rangle_c
\end{eqnarray}
where $I^0$ is the time component of the current and ${\vec Q}$
has been taken to be parallel to the $z$-axis. The charge radii
can be obtained by $\langle r^2\rangle_\pi=-6(dF_C/dQ^2)_{Q^2=0}$·

In front form, in the $Q^+=0$ frame, the charge form factor is
extracted from the ``plus'' component of the current, $I^+=n\cdot
I$, with $n=\{-1,0,0,1\}$:
\begin{eqnarray}
F_C(Q^2)=\langle0|I^+(0)|0\rangle \, ,
\end{eqnarray}
in this case the momentum transfer is taken to be transverse to
the $z$-direction~\cite{Riska04}.

For vector mesons, such as the $\rho$ and the $K^*(892)$, the
definition of Ref.~\cite{Chung} is adopted. For point and instant
forms, we have:
\begin{eqnarray}
G_C(Q^2)&=& {1\over3}\left [\langle0,\vqhalf|I^0(0)|-\vqhalf,0\rangle_c \right.\nonumber \\
&&\left.+2\langle1,\vqhalf|I^0(0)|-\vqhalf,1\rangle_c \right], \nonumber\\
G_M(Q^2)&=&\sqrt{2\over\eta}
\langle1,\vqhalf|I_+(0)|-\vqhalf,0\rangle_c, \label{eq:rhoFFpi}\\
G_D(Q^2) &=&\frac{1}{2\eta}
\left[\langle0,\vqhalf|I^0(0)|-\vqhalf,0\rangle_c \right.\nonumber \\
&&\left.-\langle
1,\vqhalf|I^0(0)|-\vqhalf,1\rangle_c \right],\nonumber
\end{eqnarray}
where $I_+=(1/2)(I_x + iI_y)$. For front form,
\begin{eqnarray}
&&G_{C}(Q^2)=F_{0d}+\osix F_{2d}
-\tthird\eta \left\{F_{0d}+F_{2d}+\fivehalf F_{1d}\right\},\nonumber\\
&&G_{M}(Q^2)=2F_{0d}+F_{2d}+F_{1d}(1-\eta),\nonumber\\
&&G_{D}(Q^2)=\frac{1}{\eta}\left\{F_{2d}+\eta\left(\ohalf
F_{2d}-{F_{0d}-F_{1d}}\right)\right\}\,, \label{eq:rhoFFf}
\end{eqnarray}
where
\begin{eqnarray}
&&F_{0d}(Q^2)=\frac{1}{2(1+\eta)}
\{\langle1|I^+(0)|1\rangle+\langle0|I^+(0)|0\rangle\},\nonumber\\
&&F_{1d}(Q^2)=\frac{-\sqrt{2}}{\sqrt{\eta}(1+\eta)}\langle1|I^+(0)|0\rangle,\nonumber\\
&&F_{2d}(Q^2)=\frac{-1}{(1+\eta)}\langle1|I^+(0)|-1\rangle \, .
\label{eq:Fi}
\end{eqnarray}
The kinematical variable $\eta$ is defined as
$\eta={1\over 4}(v_f-v_a)^2=Q^2/4M^2$, where $M$ is the meson mass.

For each form of kinematics the dynamics generates the current
density operator from a kinematic current. For point form we have,
\bea
&&\bra{\vec v_f,\vec v_2'}I^\mu(0) \ket{\vec v_2,\vec v_a}= \nonumber \\
&&\delta^{(3)}(v_2'-v_2)(\osix+\half \tau_3^{(1)}) \bar u(\vec
v_1\,') \gamma^{(1)\mu}u(\vec v_1)\, , \label{cur4}
\eea
for front form,
\beqa
&&\bra{P^+, P_{\perp f},{\bf p}_2'}I^+(x^-,x_\perp)
 \ket{{\bf p_2}, P_{\perp a},P^+}= \\
&&\delta^{(3)}(p_2'-p_2)(\osix+\half \tau_3^{(1)}) \bar u({\bf
p_1'}) \gamma^{(1)+}u({\bf p_1}) e^{\imath( P_{\perp f}- P_{\perp
a})\cdot  x_\perp}\, ,\nonumber
\eeqa
and for instant form,
\beqa
&&\bra{\half \vec Q,\vec p'_2}I^\mu(\vec x) \ket{\vec p_2,-\half \vec Q }= \\
&& \delta^{(3)}(p_2'-p_2)(\osix+\half \tau_3^{(1)}) \bar u(\vec
p_1\,')\gamma^{(1)\mu}u(\vec p_1) e^{\imath(\vec Q\cdot \vec
x)}\,.\nonumber
\eeqa

The meson decay constant can be obtained from the following matrix
element~\cite{Jaus},
\begin{eqnarray}
&&\langle 0|\bar{q_1}\gamma^\mu\gamma_5q_2|P\rangle
=iP^\mu\sqrt{2}f_P \,,\nonumber \\
&&\langle 0|\bar{q_1}\gamma^\mu
q_2|V\rangle=M_V\epsilon^\mu(p)\sqrt{2}f_V \, .
\end{eqnarray}
In front form it translates to~\cite{Cardarellidecay},
\begin{eqnarray}
f_P&=&\frac{\sqrt{6}}{(2\pi)^{3/2}}\int d\xi
d^2k_\perp\sqrt{\cal{J}}\varphi(k^2)  \\
&\times&
\frac{[(1-\xi)m_1+\xi m_2]}
{\sqrt{\xi(1-\xi)[M_0^2-(m_1-m_2)^2]}} \,, \nonumber \\
f_V&=&\frac{\sqrt{6}}{(2\pi)^{3/2}}\int d\xi
d^2k_\perp\sqrt{\cal{J}}\varphi(k^2) \nonumber \\
&\times&
\frac{[(1-\xi)m_1+\xi
m_2+\frac{2p_\perp^2}{M_0+m_1+m_2}]}
{\sqrt{\xi(1-\xi)[M_0^2-(m_1-m_2)^2]}} \,.
\end{eqnarray}
In point and instant form, the temporal component of the current
is considered, together with $\bm P=0$. Then we have $\bm
P_0\rightarrow M$, $d^3p\rightarrow d^3k$, ${\cal J}\rightarrow1$,
wave function
$R^{00}_{\lambda\bar{\lambda}}(v_{K},\bm{k})\rightarrow
\sqrt{2}\lambda\delta_{\lambda,-\bar{\lambda}}\varphi(k^2)$, where
$\lambda,\bar{\lambda}$ are the spin projection variables. Finally
it can be seen that~\cite{Ebert},
\begin{eqnarray}
f_P&=&\frac{\sqrt{3}}{(2\pi)^{3/2}\sqrt{M}}\int
d^3k\varphi(k^2) \nonumber \\
&\times&
\frac{(m_1+\omega_1)(m_2+\omega_2)-\vec{k}^2}
{2\sqrt{\omega_1\omega_2(m_1+\omega_1)(m_2+\omega_2)}} \, ,\nonumber\\
f_V&=&\frac{\sqrt{3}}{(2\pi)^{3/2}\sqrt{M}}\int
d^3k\varphi(k^2)\nonumber \\
&\times&
\frac{(m_1+\omega_1)(m_2+\omega_2)-\vec{k}^2+2\vec
k_\perp^2} {2\sqrt{\omega_1\omega_2(m_1+\omega_1)(m_2+\omega_2)}}
\, . \label{eq:dc}
\end{eqnarray}
Instant and point form share the same formula due to the $\bm P=0$
requirement.

In the nonrelativistic limit, $\vec k^2/m^2\rightarrow0$,
Eq.~(\ref{eq:dc}) becomes:
\begin{eqnarray}
f^{NR}_P=\frac{{\cal N}}{\sqrt{M_{P}}}|\varphi(0)|,\hspace{10pt}
f^{NR}_V=\frac{{\cal N}}{\sqrt{M_{V}}}|\varphi(0)| \, ,
\end{eqnarray}
giving the nonrelativistic predictions
$f_\pi/f_{\rho}=\sqrt{m_\rho/m_\pi}\approx2.3$ and
$f_K/f_{K^*(892)}=\sqrt{m_{K^*(892)}/m_K}\approx1.4 $,
these are in disagreement with the
experimental data as will be discussed in Section~\ref{sec:kst}.

\section{Numerical results of $\pi$ and $K$}
\label{sec:pik}

Employing the formalism described in Sections~\ref{sec:wf}
and~\ref{sec:ew}, the form factors, charge radii and decay
constants of $\pi$ and $K$ can be obtained in the three forms of
kinematics considered.

\subsection{Instant form results}

The pion electromagnetic form factor is well described in instant
form with the simple wave functions considered as can be seen in
Fig. 1 of Ref.~\cite{junhe}.

In Table~\ref{tab:resIform} the values for the decay constants and
charge radii of the considered mesons obtained in instant form are
presented. Although the overall agreement with the data is quite
good, there are discrepancies both in the decay constant of the
kaon, which is off by 20 \%, and in the squared charge
radius of the $K^0$ which is off by 30 \%.

Therefore with the parameters used for the light quark sector the
sizes of $K$ and $K^*(892)$ are found to be larger than the
experimental value. We can accommodate the experimental data by
using a more compact wave function for the mesons containing an
$s$ quark, larger $b$ in the gaussian case, which will decrease
the charge radii.

The use of two parameter sets, one for mesons made up of light
quarks, $\pi$ and $\rho$, and another one for mesons containing a
strange quark, $K$ and $K^*(892)$, could be due to the different
masses of the $u(d)$ and $s$ quarks or to a possible difference in
the dynamics of the $s$ quark that, in our framework, may be
accounted for by slightly changing the mass operator. These
readjusted parameters are given, when different from the original
ones, in brackets in Table~\ref{parameters}. With them, results in
brackets, the agreement with data in Table~\ref{tab:resIform}
improves specially for the decay constant of the $K$.

\begin{table*}[h]
\begin{center}
\begin{tabular}{c|c|c|c|c|c}
\hline
w.f.      &  $f_\pi$ [MeV] &  $f_K$ [MeV] &$\sqrt{\langle r^2_\pi\rangle}\ [$fm$]$ &$\langle r^2_{K^\pm}\rangle\ [$fm$^2]$
 &$\langle r^2_{K^0}\rangle\ [$fm$^2]$ \\
\hline
Rational  &   104.4        &87.0 [111.8]     &  0.619            & 0.40 [0.31]    & $-$0.115 [$-$0.055] \\
\hline
Gaussian  &   95.6         &82.4 [112.2]     &  0.600            & 0.38 [0.22]    & $-$0.110 [$-$0.050] \\
\hline\hline
Exp.      & 92.4 $\pm 0.33$&113.0 $\pm 1.3$  &0.663 $\pm 0.006$  & 0.34 $\pm 0.05$& $-$0.076$\pm$ 0.018\\
\hline\hline
\end{tabular}
\caption{Decay constants and charge radii of $\pi$ and $K$
obtained in instant form. The values, the ones within brackets,
are obtained using the parameters in Table~\ref{parameters}. The
experimental data are form
Refs.~\cite{PDG,Amendolia,Amendoliakaon}.
\label{tab:resIform}}
\end{center}
\end{table*}

\begin{figure}[h]
\vspace{20pt} \mbox{\epsfig{file=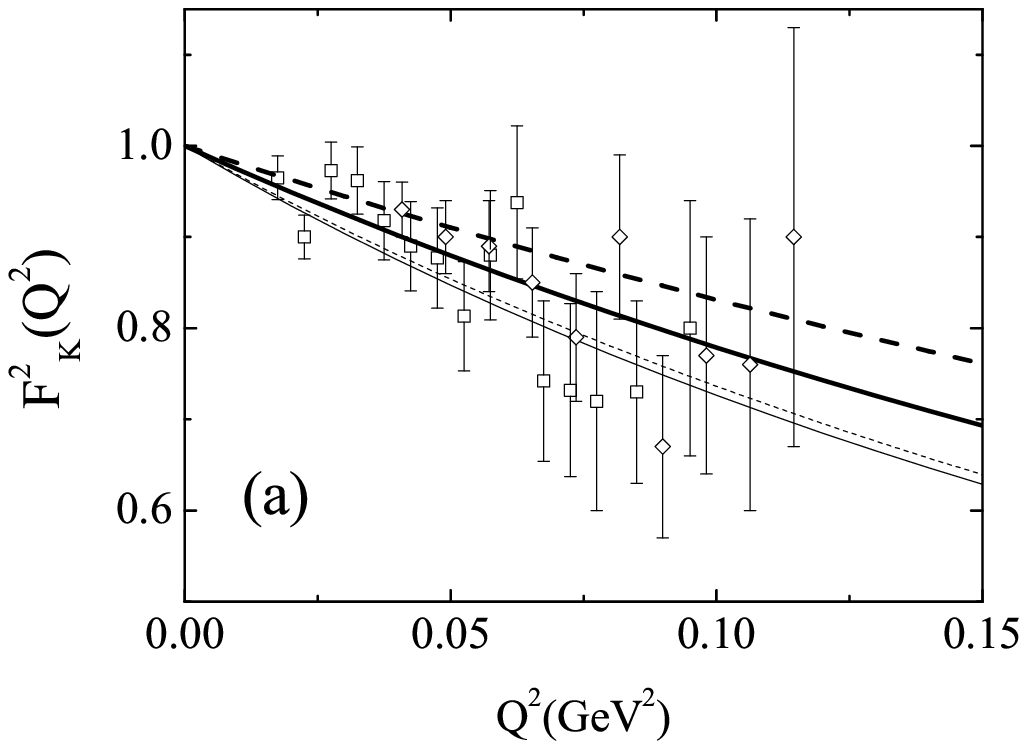, width=80mm}}
\mbox{\epsfig{file=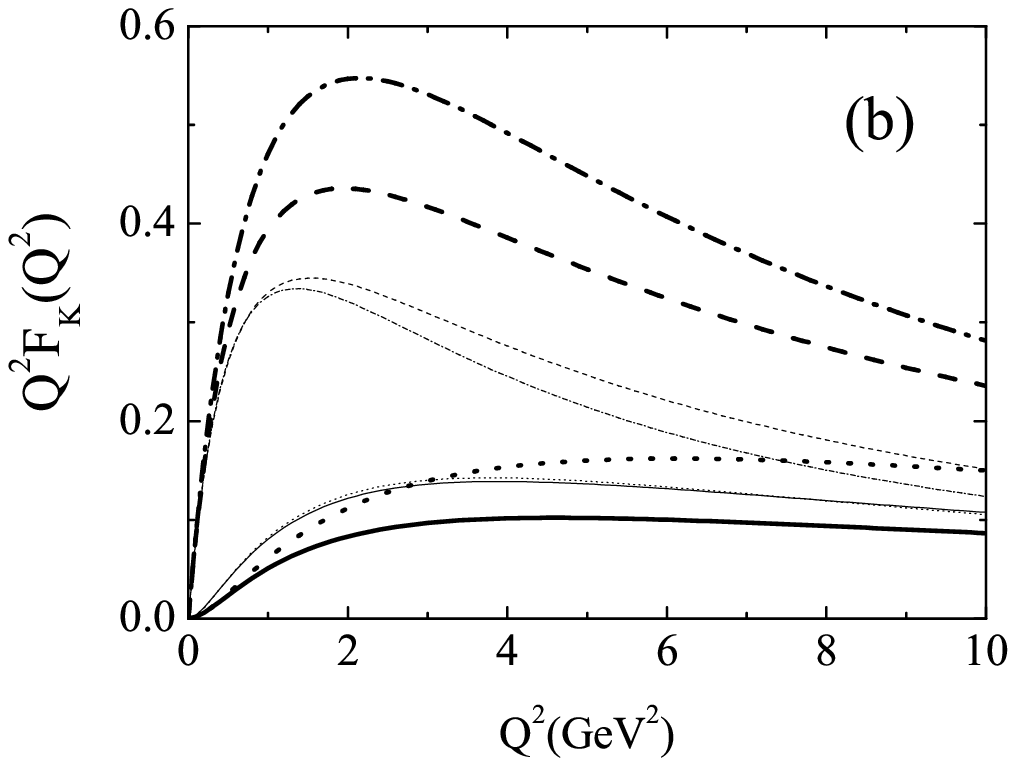, width=80mm}} \caption{ (a) $K$ charge
form factor squared in instant form as function of $Q^2$(GeV$^2$).
Solid and dashed lines stand for rational and gaussian wave
functions. (b) $K$ charge form factor in instant form multiplied
by $Q^2$. Solid and dotted lines correspond to the $K^0$ form
factor using rational and gaussian wave functions respectively.
Dashed and dot-dashed lines correspond to the $K^+$ form factor
using rational and gaussian wave functions respectively. Thin
lines are obtained with the parameters in Table~\ref{parameters}
while thick lines are obtained with the readjusted parameters
given in the same table. The experimental data are from
Refs.~\protect\cite{Amendoliakaon,Dally}. \label{fig:kI} }
\end{figure}

In Fig.~\ref{fig:kI} the obtained charge form factors of the kaon
are presented. Both using the original set of parameters as well
as with the readjusted ones, the available experimental data are
correctly reproduced. The high-$Q^2$ behavior in instant form can
thus be considered as a prediction once the parameters of the wave
function were already fixed. The high-$Q^2$ behavior is close to
$\propto 1/Q^2$ for the $K^0$ case, which is also the predicted
behavior in QCD~\cite{Farrar}. However, as already occurred with
the $\rho$ form factors~\cite{junhe}, the fall off of the form
factor of the $K^+$ is faster than $\propto 1/Q^2$ being closer to
$\propto 1/Q^4$.

\subsection{Point form results}

\label{subsec:point form}

In Ref.~\cite{junhe} the results obtained for the pion form factor
were presented explicitly emphasizing the fact that it was not
possible to find a ground state, within the considered wave
functions, that would reproduce the $Q^2$ behavior with reasonable
values for the parameters.

A first glance at Table~\ref{tab:resPform} tells us that point
form does not provide a plausible description of the experimental
data when the parameters of Table~\ref{parameters} are used.
However, as we were not able to constrain our parameters with the
$\pi$ data, here a different approach will be followed.
Considering that the $\pi$ might be a pathology, maybe not a
simple $q\bar{q}$, we choose to concentrate on the ability of the
point form approach to describe the data of the $K$.

\begin{table*}[h]
\begin{center} \begin{tabular}{c|c|c|c|c|c}
\hline
\multicolumn{6}{c}{Point Form}\\
\hline
        &  $f_\pi$        &  $f_K$               &$\sqrt{\langle r^2_\pi\rangle}\ [$fm$]$
                         &$\langle r^2_{K^\pm}\rangle\ [$fm$^2]$&$\langle r^2_{K^0}\rangle\ [$fm$^2]$ \\
\hline
Rational& 9730.1 [104.4]    &6902.6 [111.8]   & 2.55 [3.12]         & 0.52 [1.66]      & $-$0.003 [$-$0.477] \\
\hline
Gaussian& 838.5 [95.6]      &512.1 [112.2]    & 3.02 [3.23]         & 0.74 [1.50]      & $-$0.008 [$-$0.411] \\
\hline\hline
Exp.    &   92.4 $\pm 0.33$ &113.0 $\pm 1.3$  &  0.663 $\pm 0.006$  & 0.34 $\pm 0.05$  & $-$0.076$\pm$ 0.018\\
\hline\hline
\end{tabular}
\caption{Decay constants and charge radii of $\pi$ and $K$ in
point form. Same description as Table.~\ref{tab:resIform}.
\label{tab:resPform}}
\end{center}
\end{table*}

Due to the fact that the decay constant is calculated with the
same formula in both instant and point forms, see Eq.~(\ref{eq:dc}),
we decided to use the instant form values to
get a closer description of the data in our point form calculation.
The results are also shown, within brackets, in Table~\ref{tab:resPform}.
In this case the agreement of decay constants improves while the
charge radii become badly overestimated. The overestimated charge
radii can be traced back to the dependence of the form factor on
the momentum transfer $Q$ through the velocity of the system in
the Breit frame which involves the ratio $Q/(2M)$~\cite{Desplanques0407074}.

\begin{figure}[t]
\vspace{20pt} \mbox{\epsfig{file=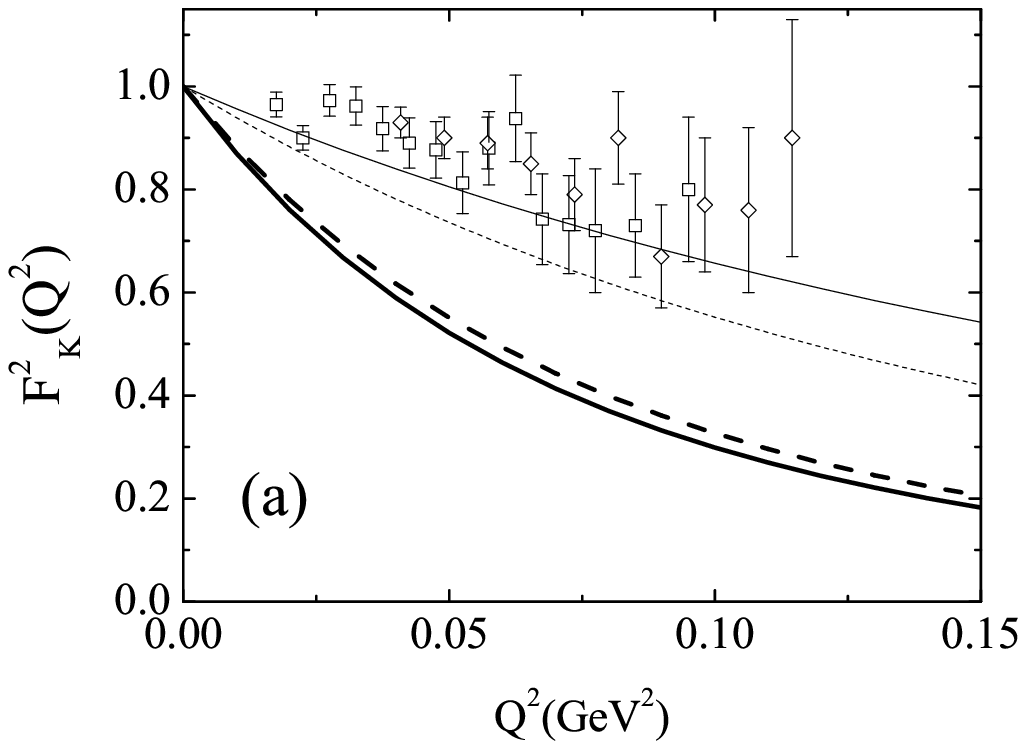, width=80mm}}
\mbox{\epsfig{file=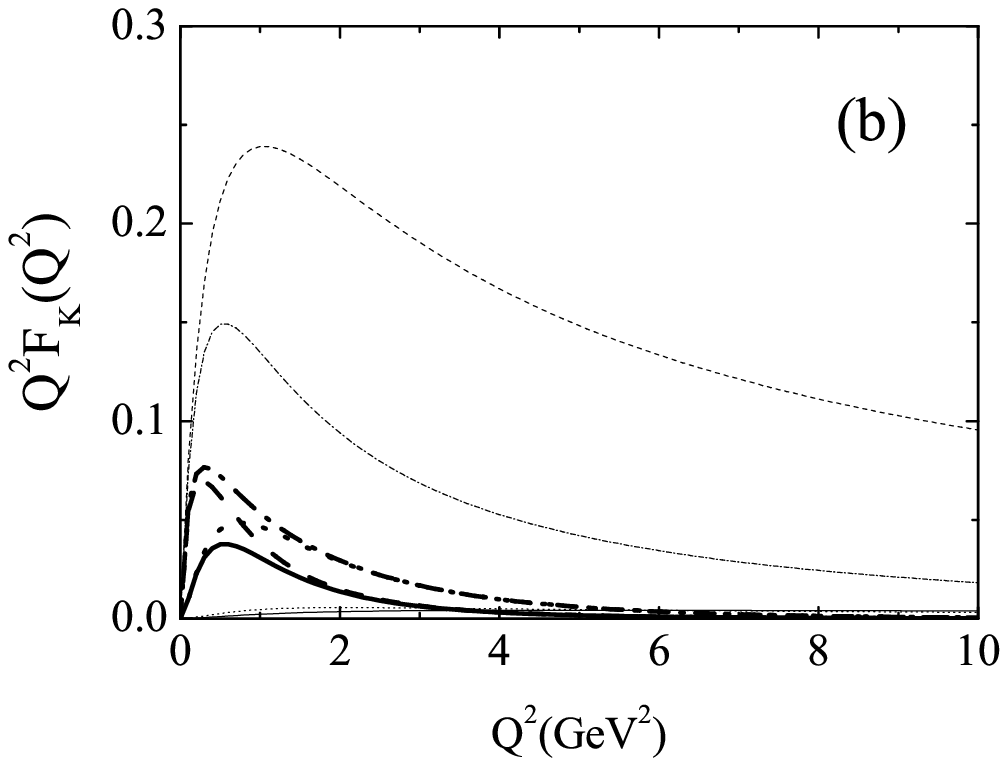, width=80mm}}
\caption{(a) $K$ charge form factor squared in point form as function of
$Q^2$(GeV$^2$). (b) $K$ charge form factor in point form
multiplied by $Q^2$. Same description as Fig.~\ref{fig:kI}.
\label{fig:kP}}
\end{figure}

Fig.~\ref{fig:kP} depicts the $Q^2$ behavior of the form factor
obtained in point form. The obtained kaon form factor in point
form is neither completely off as occurred with the pion nor
similar to the results with the other two forms as in the case of
the rho. It suggests that with mesons of increasing mass, the form
factor improves. This is consistent with one of the conclusions in
Ref~\cite{junhe} where it is pointed out that the failure of point
form to reproduce the $\pi$ form factor is most likely due to its
small mass. Indeed, a small mass indicates large effects due to
interactions which we partially neglect when considering the single quark
current approximation. Therefore, single quark currents are not
enough in point form for low mass mesons.

\subsection{Front form}

The pion electromagnetic form factor is well described in front
form with the simple wave functions considered as can be seen in
Fig. 1 of Ref.~\cite{junhe}. The decay constants and charge radii
of $\pi$ and $K$ in front form are given in Table~\ref{tab:decF}.
The first relevant result is that, with the same ground
state wave function that permitted a description of the $\pi$ and
$\rho$ form factors, reasonable values for the decay constants and
charge radii of the $K$ are obtained. The disagreement with
experimental data is less than 10 $\%$ in the decay constants and
charge radii with both shapes of the wave function. The overall
agreement with experimental data can be slightly improved by
considering a little larger size parameter, $b$ from 600 MeV to
650 MeV in the rational case and from 450 MeV to 500 MeV in the
gaussian case, as is shown in brackets in the table.

\begin{table*}[h]
\begin{center}
\begin{tabular}{c|c|c|c|c|c}
\hline \hline w.f.           &  $f_\pi$ [MeV] &$f_K$ [MeV]
&$\sqrt{\langle r^2_\pi\rangle}\ [$fm$]$
                         &$\langle r^2_{K^\pm}\rangle\ [$fm$^2]$&$\langle r^2_{K^0}\rangle\ [$fm$^2]$  \\
\hline
Rational   &   98.6 [102.5]    &114.0 [119.4]   &  0.659 [0.679]      & 0.43 [0.38]     & $-$0.080 [$-$0.069] \\
\hline
Gaussian   &   92.2 [97.2]     &106.2 [113.0]   &  0.665 [0.630]      & 0.43 [0.36]     & $-$0.077 [$-$0.062] \\
\hline
Exp.       &   92.4 $\pm 0.33$ &113.0 $\pm 1.3$ &  0.663 $\pm 0.006$  & 0.34 $\pm 0.05$ &$-$0.076$\pm$ 0.018\\
\hline \hline
\end{tabular}
\caption{Decay constants and charge radii of $\pi$ and $K$ mesons
in front form. Same description as Table.~\ref{tab:resIform}.
\label{tab:decF}}
\end{center}
\end{table*}

\begin{figure}[h]
\vspace{20pt} \mbox{\epsfig{file=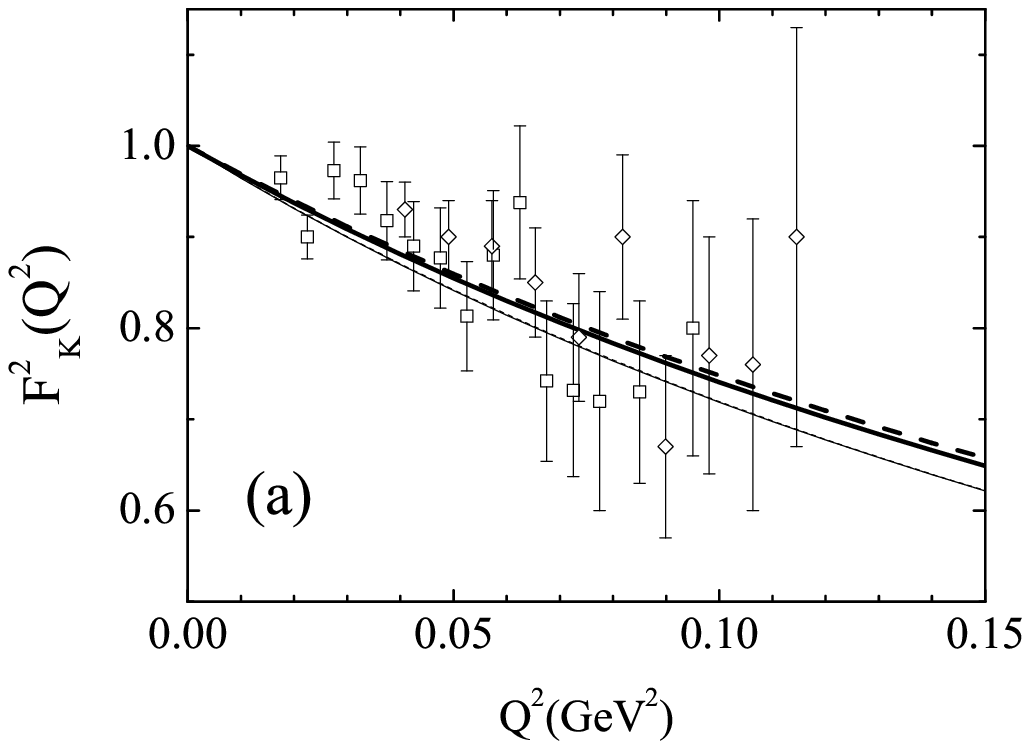, width=80mm}}
\mbox{\epsfig{file=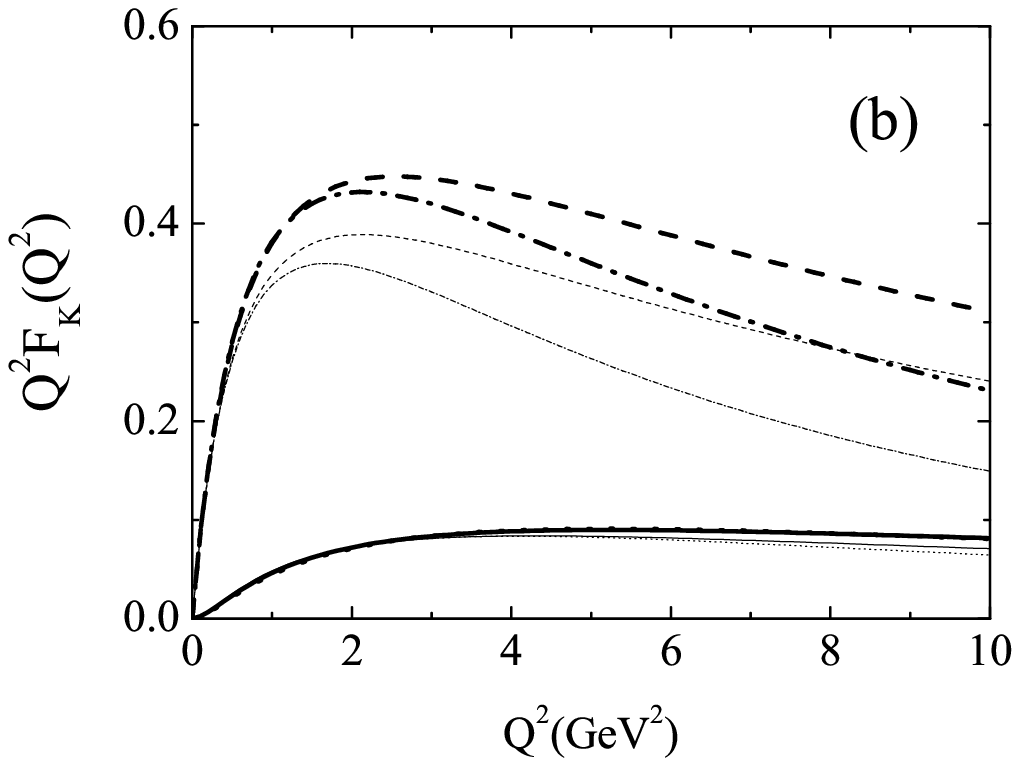, width=80mm}}
 \caption{
(a) $K$ charge form factor squared in front form as function of
$Q^2$(GeV$^2$). (b) $K$ charge form factor multiplied by $Q^2$.
Same description as Fig.~\ref{fig:kI}.  \label{fig:kF}}
\end{figure}

The results with the two sets of parameters are similar and the
available experimental data are correctly reproduced. The
high-$Q^2$ behavior can thus be considered as the front form
prediction once the parameters of the wave function were already
fixed. These results are similar to other front-form
results~\cite{Cardarelli,ChoiKaon}. The high-$Q^2$ behavior is
close to $\propto 1/Q^2$ for the $K^0$ case as occurred in the
instant form case. However, as happened with the $\rho$ form
factors~\cite{junhe} the fall off of the $K^+$ is faster than
$\propto 1/Q^2$.

\section{Form factors and decay constant of the $K^*(892)$}

\label{sec:kst}

The decay constant and electromagnetic form factor of the
$K^*(892)$ are presented using the three forms of kinematics.

In Table~\ref{tab:rhoKF} the decay constant of $\rho$ and
$K^{*}(892)$ are presented. They have been obtained using the
parameters in Table~\ref{parameters}, the ones in brackets have
been obtained with the readjusted parameters.

\begin{table*}[h]
\begin{center}
\begin{tabular}{c|c|r|r|r}
\hline \hline
             & wave function  &     $f_{\rho}$ [MeV] & $f_{K^*(892)}$ [MeV]  & $f_{K^*(892)}/f_{\rho}$  \\
\hline \hline
Instant Form & Gaussian  &   88.2[128.2]       & 97.4 [125.3]      & 1.10[0.98]               \\
             & Rational  &   96.1[129.1]       & 94.9 [124.5]      & 0.99[0.96]              \\
\hline
Point Form   & Gaussian  &   1842.1            & 1727.4            & 0.94              \\
             & Rational  &   5.952$\times 10^6$& 5.505$\times 10^6$& 0.92              \\
\hline
Front Form   & Gaussian  &  151.3 [168.3]      & 153.1 [170.5]     & 1.01 [1.01]        \\
             & Rational  &  175.4 [190.1]      & 177.6 [192.6]     & 1.01 [1.01]        \\
\hline
EXP          &           &   152.8             &  159.3            &  1.04              \\
\hline \hline
\end{tabular}
\caption{Decay constants and charge radii of $\rho$ and $K^*(892)$
mesons in instant, point and front forms. The point form values
using the readjusted parameters are the same as the instant form
values with the same parameters.\label{tab:rhoKF}}
\end{center}
\end{table*}

First, we note that with all the forms $f_{K^*(892)}/f_\rho \approx 1$
regardless of the values obtained for each of the decay constants.
In particular, the point form values are badly wrong with the
original set of parameters, similarly to what was observed in
Table~\ref{tab:resPform} for the $\pi$ and $\rho$. If the readjusted
set of parameters is used in point form the obtained values would
be the same as the instant form ones.

The values quoted as experimental in the Table are taken from
Refs.~\cite{Ivanov,Maris}. The decay constant, $g_\rho$, which
enters in the process $\rho^0\rightarrow e^+e^-$, is obtained
from the matrix element:

\beq
\frac{M_\rho^2}{g_\rho}\epsilon^\lambda_\mu(p)=\langle0|\bar{u}\gamma_\mu
u|\rho^0_\lambda(p)\rangle=\frac{1}{\sqrt{2}}\langle0|\bar{u}\gamma_\mu
d|\rho^-_\lambda(p)\rangle\,, \label{eq:grho} \eeq so that \beq
f_\rho=\frac{M_\rho}{g_\rho} \, .
\label{eq:frho grho}
\eeq

The experimental decay width for that channel, 6.77 keV~\cite{PDG},
leads to the value $g_\rho=5.03$, that corresponds to $f_\rho$=152.8 MeV.

From the partial decay widths of the processes $\tau \to V \nu_\tau$
Maris {\it et al.} extract the ratio $f_{K^*(892)}/f_\rho=1.04$. Which
implies $f_{K*}=159.3$ MeV (which is comparable to the result of
Ref.~\cite{Jaus}, $f_{K*}=153$ MeV).

The instant form calculation (with both the original and the
readjusted parameters) underestimates the decay constants by at
least 20$\%$. However, the predicted ratios $f_\pi/f_\rho$ and
$f_K/f_{K^*(892)}$ are in better agreement with experimental data
than their nonrelativistic counterparts which are given at the end
of Sect~\ref{sec:ew}.

The front form results are in better agreement with the values
extracted from experiment specially for the gaussian case. The
readjusted parameters do not improve the results in this case.

\begin{figure}[t]
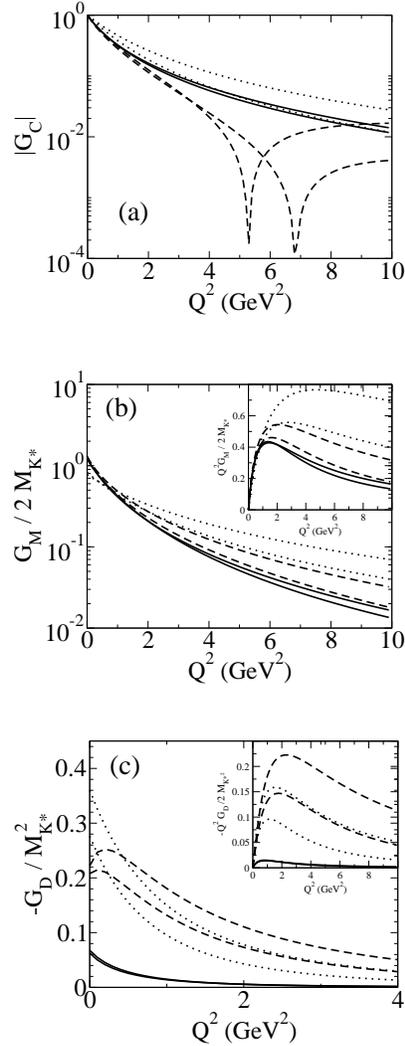

\mbox{\epsfig{file=fig4a, width=52mm}}
\vspace{20pt}

\mbox{\epsfig{file=fig4b, width=52mm}}
\vspace{20pt}

\mbox{\epsfig{file=fig4c, width=52mm}}
 \caption{Electromagnetic form factor of the $K^*(892)$
obtained in the different forms of kinematics using gaussian wave
functions. Solid, dotted and dashed lines stand for instant, point
and front form of relativistic kinematics. (a), (b) and (c) show
the coulomb, magnetic and dipole form factors, respectively.
\label{fig:kS}}
\end{figure}

The electromagnetic form factors of the $K^*(892)$ have also been
evaluated. In Fig.~\ref{fig:kS} the coulomb, magnetic and dipole
form factors of the $K^{*+}(892)$ are shown calculated making use of
Eqs.~(\ref{eq:rhoFFpi}) in the three different forms with the
parameters in Table~\ref{parameters}.

Qualitatively the obtained form factors are quite similar to the
$\rho$ form factors presented in~\cite{junhe}. In fact, as was
already reported for the nucleon form factors~\cite{Riska04} and
also for the charge form factor of the $\rho$, a node is found in
the charge form factor of the $K^{*+}(892)$ in front form. The
node is in this case at $Q^2$ around 6 GeV$^2$. The predicted
behavior for $G_M$ is similar in instant and front forms, being
considerably smaller in point form.

\begin{figure}[t]
\vspace{20pt} \mbox{\epsfig{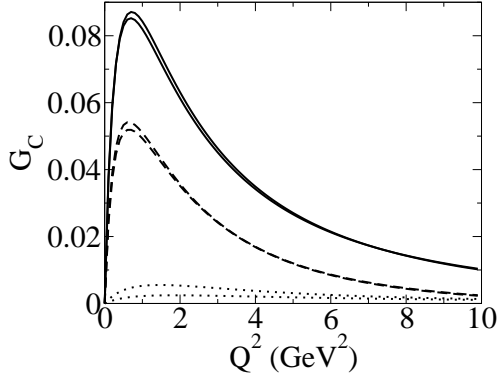}}
\caption{Coulomb form factor of the $K^{*0}(892)$
obtained in the different forms of kinematics using gaussian wave
functions. Solid, dotted and dashed lines stand for instant, point
and front form of relativistic kinematics.
\label{fig:kS0}}
\end{figure}

The relativistic nature of the calculation produces non-zero
values for the charge form factors of the $K^{*0}(892)$ even
in the $SU(3)_f$ symmetric case. However, the relativistic
effect alone is much smaller than the effect arising from the
actual existing mass difference between the $s$ and $u(d)$ quarks.
In Fig.~\ref{fig:kS0} the charge form factor of the $K^{*0}(892)$ is
presented.

\begin{figure}[t]
\vspace{20pt} \mbox{\epsfig{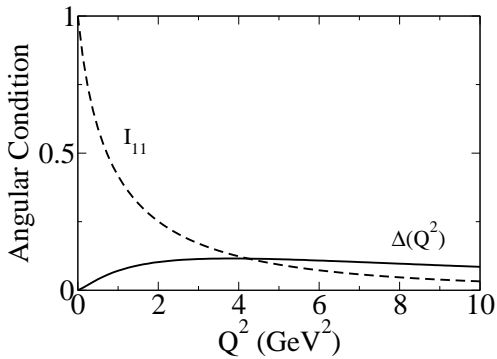}}
 \caption{In solid the angular condition defined in Eq.~(\ref{eq:ac}),
dashed corresponds to the matrix element
$\langle1|I^+(0)|1\rangle$. Both for the case of the $K^*(892)$.
\label{fig:ac}}
\end{figure}

Finally, as is well known~\cite{Coester92,Keister} the use of
single quark currents does not permit an unambiguous extraction of
the form factors in front form from the considered matrix
elements. In this work the form factors are extracted from the
same matrix elements as used in Ref.~\cite{Chung} for the case of
the deuteron. There, due partly to the large mass of the deuteron
as compared to that of the constituents, the breaking in
rotational symmetry due to the fact that single quark currents
were employed was small. This can be estimated by showing what is
called the ``angular condition''. This is a certain linear
combination of the four matrix elements used to extract the form
factors which should vanish would the calculation be rotationally
invariant. It can be defined as,
\beq
\Delta(Q^2) = (1+2\eta)I_{1,1} + I_{1,-1} -\sqrt{8\eta} I_{1,0} -I_{0,0} \, .
\label{eq:ac}
\eeq
In Fig.~\ref{fig:ac} the function $\Delta(Q^2)$ is shown as function
of $Q^2$ together with the matrix element $I_{1,1}$ which should
serve to compare the magnitude of the angular condition. This figure shows
that above a certain $Q^2$ the obtained
behavior of the form factors of spin 1 systems in front form using
single quark currents should be taken with care.

\section{Discussion and summary}

In the quark model the main difference between ($\pi$,$\rho$) and
($K$,$K^*(892)$) is the substitution of a light quark by a strange
quark in the latter. Thus it is natural to explore the
($K$,$K^*(892)$) system using the mass operator, which was
originally fixed to reproduce the $\pi$ charge form factor, as
starting point.

We have presented charge form factors, charge radii and decay
constants of the $\pi$, $K$ and $K^*(892)$ making use of the
three different forms of relativistic kinematics.

In front form, in the impulse approximation and with simple wave
functions, the charge form factors, charge radii and decay
constants of the $\pi$, $K$ and $K^*(892)$ are reasonably
reproduced making use of the same parameters that were employed
previously to study the $\rho$ form factors. An effective way
to phenomenologically account for the differences arising
from the presence of a $s$ quark would be to allow for a
variation of the mass operator. This is achieved by considering
a slightly different set of parameters for the ($K$, $K^*$)
system. In the front form case, however, the agreement
with the data is already acceptable with the original set.

In instant form a slightly worse description of the data is
achieved, specially in the case of the vector meson decay
constants which are underestimated by 30 $\%$. An improved
description of the data can be achieved if two different sets
of parameters, one for the mesons made up of light quarks,
$\pi$ and $\rho$, and another one for the mesons which contain
a strange quark, $K$ and $K^*(892)$, is employed. The
readjusted values which essentially correspond to a more
compact wave function in coordinate space imply a smaller
radius for the systems containing an $s$ quark.

The description of the data using point form is very poor.

The approach followed here, used also in Ref.~\cite{junhe}, tries
to assess to what extend the existing data for meson form factors
can be reproduced using simple assumptions for the mass operator
and current operators using the different forms of relativistic
quantum mechanics. The results presented here together with other
recent ones, e.g., \cite{Desplanques0407074,Riska04} show that the
standard realization of the front form, $Q^+=0$, tends to give
results which are closer to experimental data. The instant form
results are also qualitatively similar. Thus, one can legitimately
wonder why the point form fails. Ref.~\cite{Desplanques0407074}
deals with this very question and finds that a common feature
which is shared by the ``successful'' implementations is the fact
that in all cases momenta is conserved at the interaction vertex.
The point form, and also front form in the $Q^+\neq 0$ frame, do
not conserve momenta at the quark interaction vertex, which could
indicate that the requirement of translation invariance at the
quark level would be a much more relevant one. Our work does not
contradict those lines.

The same prediction for the high-$Q^2$ behavior of the form factor
of the $K^0$ is found with both instant and front forms. The
predicted behavior is close to the QCD prediction of Refs.~\cite{Farrar}.
For the charged kaon the high-$Q^2$ behavior of the form factors is
closer to $\propto 1/Q^4$ that to $1/Q^2$ both in instant and front
forms. The disagreement with the asymptotic QCD behavior may be due
to the simple assumptions for the electromagnetic current or
simply to the fact that pQCD is not reached as such momentum
transfers in this specific problem. In the front form case
some care must be taken when considering the form factors of spin-1 mesons.
The ambiguity arising when working with single quark currents in
the definition of the form factors becomes relevant for low mass
systems.

The ratio $f_{\pi}/f_\rho$ is considerably improved when
relativity is taken into account, as compared to the
nonrelativistic results. Although the values for the decay
constants for $K^*(892)$ are not close to the experimental data in all
the forms, it is found that the $f_{K^*(892)}/f_\rho$ is well
reproduced with each of them.

Similar to what was found when studying the form factors of the
nucleon and the $\rho$, the charge form factor of the $K^*(892)$
in front form contains a node close to 6 GeV$^2$. This node could,
unlike the one in the nucleon electric form factor, disappear when
two body currents are incorporated in the framework.

\begin{acknowledgement}
The authors want to thank D. O. Riska for valuable comments on the
manuscript. This work is supported by the National Natural Science
Foundation of China (No. 10075056 and No. 90103020), by CAS
Knowledge Innovation Project No. KC2-SW-N02. B. J.-D. thanks the
European Euridice network for support (HPRN-CT-2002-00311), the
Academy of Finland through grant 54038 and the National Science
Foundation, grant No. 0244526 at the University of Pittsburgh.
\end{acknowledgement}

\end{document}